\documentstyle[12pt]{article}
\newcommand{\nc}{\newcommand}
\nc{\beq}{\ \\[3mm]\bdm\refstepcounter{equation}\hff}
\nc{\eeq}{\hff(\theequation)$\\[3mm]}
\nc{\bed}{\ \\[3mm]\bdm{}\hff}
\nc{\eed}{{}\hff$\\[3mm]}
\nc{\preprint}[1]{\begin{table}[t]\begin{flushright}
    \begin{large}{#1}\end{large}\end{flushright}\end{table}}
\nc{\eref}[1]{(\ref{#1})}
\nc{\prt}{\partial}
\nc{\cb}{\cr\noalign{\vspace{3pt}}}
\nc{\DD}[2]{\frac{d^{#2}}{d#1^{#2}}}
\nc{\PD}[2]{\frac{\prt^{#2}}{\prt #1^{#2}}}
\nc{\bdm}{$ \displaystyle}
\nc{\EQ}[1]{\hff\prqn\label{#1}\eie}
\nc{\bie}{\ $\\[3mm]{} \displaystyle}
\nc{\nie}{\\[1mm]{}\displaystyle}
\nc{\prqn}{\refstepcounter{equation}\hf (\theequation)}
\nc{\eie}{$\\[3mm]}
\nc{\bea}{\ \\[3mm]\bdm\begin{array}{l}\displaystyle}
\nc{\eea}{\end{array} \hff
           \begin{array}{r}\prqn\end{array}\\[3mm]$}
\nc{\hsp}{\hspace*{1.7mm}}
\nc{\hfc}{\hf , \hf}
\nc{\hf}{\hfill}
\nc{\hff}{\hf\hf}
\nc{\al}{\alpha}
\nc{\upa}{\uparrow}
\nc{\dna}{\downarrow}
\nc{\dS}{d^3\!y}
\nc{\dV}{d^4\!x}
\nc{\tpo}{$2+1$\ }
\nc{\Tpo}{$3+1$\ }
\nc{\mdg}{\sqrt{-\det g}}
\nc{\mdG}{\sqrt{-\det G}}
\nc{\Ue}{$U(1)_{\rm e.m.}$}
\nc{\Ug}{$U(1)_{\mg}$}
\nc{\g}{\mbox{g}}
\nc{\A}{\mbox{\tt A}}
\nc{\mg}{{\rm g}}
\nc{\prd}[3]{\ { \it Phys. Rev.} {\bf D#1} (#2) #3}
\nc{\prl}[3]{\ { \it Phys. Rev. Lett.} {\bf #1} (#2) #3}
\nc{\plb}[3]{\ {\it Phys. Lett.} {\bf #1B} (#2) #3}
\nc{\npb}[3]{\ {\it Nucl. Phys.} {\bf B#1} (#2) #3}
\nc{\prs}[3]{\ {\it Proc. Roy. Soc.} {\bf #1} (#2) #3}
\catcode`\@=11

\def\section{\@startsection {section}{1}{\z@}{-13ex plus -1ex
                      minus -.2ex}{3ex plus .2ex}{\large\bf}}
\load{\scriptsize}{\rm}
\begin{document}
\preprint{BGU--10--92\\ hep-th/9302081}
\title{\Large\bf Threefold Family of Charged Spin--1/2 Dirac Bubbles}
\author{\\ {\bf Aharon Davidson\thanks{E-mail: davidson@bguvms}}
\ \ \addtocounter{footnote}{5}
and \ {\bf Uzi Paz\thanks{E-mail: uzipaz@bgumail.bgu.ac.il}} \\ \ }
\date{Physics Department \\ Ben-Gurion University of the Negev \\
Beer-Sheva 84105, Israel\\[7mm] {\footnotesize (Phys. Lett.
B, in press)}}
\maketitle

\begin{abstract}
We extend Dirac's `extensible model of the electron' to include spin and
family. \Ue\ charge conservation on the bubble is translated into a
secondary \Ug\ world-manifold gauge principle. Reflecting the secondary
magnetic monopole configuration on spatial $S^2$, the harmonic
excitations may furnish half integer $SU(2)_{\rm spin}\!\otimes\!
U(1)_{\mg}$ representations.  Our spin--$1/2$ `electron' is described by
\underline{four} world-manifold scalar fields. Its \underline{three}
varieties are associated with different minima of the ($6^{th}$--order)
surface-tension scalar potential.
\end{abstract}

\section{Introduction}
The long-lived idea of a finite size elementary particle is as appealing
as ever. Adopting Dirac's viewpoint, such an idea may be ``the most
natural concept that makes the total energy of the Coulomb field of the
electron finite". Experimentally, however, up to a scale of
$10^{-16}$cm, quarks and leptons exhibit a point-like behavior.
Nonetheless, their apparently superfluous proliferation (exactly three
families) gives us some reason to suspect an internal structure. \ The
finite electron idea has been introduced [1], on a classical level, by
Abraham and Lorentz at the dawn of special relativity. Later on,
Einstein [2] has examined the possibility that gravitational fields play
an essential role in the game. He was even ready to modify his field
equations in order to ``elaborate a theory which will account for the
equilibrium of the electricity constituting the electron". His overall
conclusion was, however, that the problem cannot be explained solely by
means of a locally varying cosmological constant.

The constitution of the elementary quanta was retackled and
sophistically refined by Dirac [3]. He pictured the electron as a
breathing bubble in the electromagnetic field, with no constraints
fixing its size and shape. An effective surface tension was postulated
in order to prevent the electron from flying apart under the Coulomb
repulsion of its surface charge. The naive, yet pedagogical `extensible
model of the electron' was an attempt to deal with Rabi's immortal
question `who ordered the muon?'. The major drawback of the model was
its impotency to incorporate spin. Nowadays, being exposed to modern
theories such as grand-unification, supergravity, superstrings, and
hypothetical quantum gravity, claiming to take us beyond the standard
Glashow-Weinberg-Salam electronuclear model, it is quite ironic that one
still lacks a decent clue regarding the family puzzle and the structure
of the Fermi mass matrix.  Imaginative preon models [4], accompanied by
hyper-color theories [5] especially designed to handle QCD-like
quark/lepton compositeness, have not shed light on these problems. The
intuitive approach advertised by Dirac, which in our opinion has not
been examined to its full depth, may eventually turn out to be a step in
the right direction.

The physics that takes us beyond the Dirac limit is formulated here in
terms of a world-manifold local gauge theory. Invoking various field and
group theoretical ideas in \tpo dimensions (i.e. on the bubble), while
fully respecting \Ue\ gauge invariance in \Tpo dimensions (i.e.
surrounding the bubble), we construct a self-consistent Lagrangian
formalism which, up to weak interactions (to be hopefully incorporated
in future stage), fairly represents a finite-size electron-like extended
object. Our work attempts to ({\bf i}) account for the spontaneous
(rather than explicit) generation of spin, ({\bf ii}) construct an
explicit spin-$1/2$ static soliton configuration in space-time, ({\bf
iii}) offer a possible solution to the family puzzle, and ({\bf iv})
re-examine the problem of mass. \ \ The formalism is P/C/T conserving,
and has nothing to do with the Chern-Simons theory, the popular source
of fractional statistics [6] in \tpo dimensions. Due to length
limitations, however, we skip some lengthy calculations (to be published
elsewhere [7]). For the sake of clarity, we proceed in steps towards
eqs.  \eref{18}, \eref{20}.

\section{U(1)$_{\rm e.m.}$ Gauge Invariance}
Our starting point is conventional. We work in a flat \Tpo dimensional
background with metric $G_{\mu\nu}(x)$. The bubble's (\tpo dimensional)
world-manifold is parametrized by $x^{\mu}(y^{\al})$, so that its
induced metric and projective electromagnetic field take the form
\bed
g_{\al\beta} \equiv G_{\mu\nu}x^{\mu}\!,_{\al}x^{\nu}\!,_{\beta} \hfc
A_{\al}\equiv A_{\mu}x^{\mu}\!,_{\al}\hsp,
\eed
respectively. We are after a self-consistent $A^{\mu}(x)$ which arises
from some given (to be dynamical) charge/current surface distribution
$Q^{\al}(y)$. The interaction of the electromagnetic field with our
bubble is conventionally described by the \Ue\ invariant action
\beq
S^{(1)}=\int\frac{1}{4}G^{\mu\lambda}G^{\nu\sigma}F_{\mu\nu}F_{\lambda
\sigma}\mdG\dV +\oint Q^{\al}A_{\al}\dS\hsp.
\eeq
Note that it is \bdm q^{\al}=\frac{Q^{\al}}{\mdg}$, and not \bdm
Q^{\al}$, which transforms like a legitimate \tpo vector.

A variation with respect to $A_{\mu}(x)$, with the bubble treated as an
internal boundary, gives rise to the boundary condition
\bed
F^{\mu\nu}n_{\nu}=\frac{Q^{\al}x^{\mu}\!,_{\al}}{\mdg}
\eed
for the free Maxwell equations outside the bubble. $n^{\mu}$ denotes
here a space-like unit normal vector pointing inwards. For the radially
symmetric case, defined by $\{ t=\tau$, $r=r(\tau)$, $\theta=\theta$,
$\phi=\phi \}$, the Coulomb potential \bdm A_t=\frac{q}{r}$ originates
from
\beq
Q^{\tau}=q\sin \theta\hfc Q^{\theta}=Q^{\phi}=0\hsp.
\eeq
Notice that in the general case, \Ue\ gauge invariance is not
automatically guarantied. In fact, for $S^{(1)}$ to stay trivial if
$A_{\mu}=\prt_{\mu}\Lambda$ (translated into $A_{\al}=
\prt_{\al}\Lambda$),\
$Q^{\al}$ better obey $Q^{\al},_{\al}=0$. For a point particle, a
constant $Q^{\tau}$ will do, but for a higher-dimensional extended
object, such a conservation law requires pre-arrangement.

\section{A Secondary Gauge Principle}
{}$Q^{\al}$ is locally conserved on the world-manifold if and only if
\beq
Q^{\al}=\epsilon^{\al\beta\gamma}\prt_{\beta}a_{\gamma}\hsp.
\eeq
The fact that $a_{\al}(y)$, which has not been explicitly introduced in
the Dirac theory, is defined up to
$a_{\al}\rightarrow a_{\al}+\lambda,_{\al}$,
need not be regarded as a mere coincidence. On the contrary, it is
highly suggestive to add the kinetic term [8]
\beq
S^{(2)}=\oint\frac{1}{4m}g^{\al\gamma}g^{\beta\delta}
f_{\al\beta}f_{\gamma\delta}\mdg\dS\hsp,
\eeq
and treat $a_{\al}(y)$ as a canonical world-manifold gauge field. The
mass scale $m$ serves here to scale $a_{\al}$ to conveniently match
the units of $A_{\mu}$.

Spherical symmetry uniquely means [9]
\beq
a_{\tau}=a_{\theta}=0\hfc a_{\phi}=\g(1-\cos\theta)\hsp,
\eeq
which is recognized as a non-singular (for a non-collapsing bubble) \tpo
dimensional magnetic-monopole-like configuration. {\sf Notice that} g,
{\sf the secondary magnetic charge, plays the double role of being the
primary electric charge $q$ as well}. \ Special attention is devoted
to the so-called Dirac pole, the trace of a hypothetical Dirac-string
[10]. In the absence of a Chern-Simons term (which directly converts
flux points to electric charges) the Dirac pole is \underline{not} an
anyon [6]. Its location at the south pole is just a matter of
convenience, nothing but choosing the $z$-axis without spoiling the
spherical symmetry. As could have been anticipated, the Dirac-pole turns
out to be an artifact only for properly quantized \Ug\ charges, and only
then, can be gauged away by means of a secondary Wu-Yang-like
construction.

An important observation is that the above monopole solution of
$S^{(1)}+S^{(2)}$ stays stable against variations with respect to
$a_{\al}(y)$, i.e. it automatically satisfies the equation of motion
\bed
\left(\frac{1}{m}\mdg
f^{\al\beta}\right)_{\!\mbox{,}\beta}=
\frac{1}{2}\epsilon^{\al\beta\gamma}F_{\beta\gamma}\hsp,
\eed
for arbitrary g. \ \ The general integration of this equation
demonstrates how our bubble acts as a superconductor [11]. One can
verify that, as far as the real photon is concerned, the combined effect
of the two $a_{\al}$ terms in the Lagrangian is identical to that of a
single mass term \bdm \frac{1}{2}m\mdg g^{\al\beta}A_{\al}A_{\beta}$.

\section{Dynamical Bubble}
Following Dirac, there should be no constraints fixing the size and
shape of the bubble. In other words, $x^{\mu}(y^{\al})$ must serve as a
canonical variable. The action integral $\oint L\dS$ is then handled in
the standard way, whereas $\int\frac{1}{4}F^2\dV$ poses a minor
technical problem when noticing, that a variation with respect to
$x^{\mu}(y^{\al})$ changes its internal boundary. Taking into account
the later effect, the corresponding equation of motion reads
\bed
\frac{\delta L}{\delta x^{\mu}}-\frac{\prt}{\prt y^{\al}}
\left(\frac{\delta L}{\delta
x^{\mu}\!,_{\al}}\right) -\frac{1}{4}\mdg G^{\nu\sigma}G^{\lambda\xi}
F_{\nu\lambda}F_{\sigma\xi}n_{\mu}=0\hsp,
\eed
with the unit normal vector $n^{\mu}$ defined earlier. The last term
expresses the back reaction, i.e. the force exerted on the bubble by
the surrounding electromagnetic field.

Unfortunately, $S^{(1)}+S^{(2)}$ is not capable of preventing the bubble
from flying apart under the Coulomb repulsion. Dirac, not exercising the
option of introducing world-manifold fields, has invoked some effective
surface-tension (positive world-manifold cosmological term)
\beq
\bar{S}^{(3)}=\oint\Lambda\mdg\dS\hsp.
\eeq
Such a description, although qualitative, is sufficient for deriving a
bounded expression for the total energy. In particular, for the
spherically-symmetric case, one obtains
\bed
E(r)=\Lambda r^2 + \frac{\g^2}{2mr^2} + \frac{3q^2}{2r}\hsp,
\eed
where the electromagnetic surface contribution \bdm\frac{q^2}{r}$
appears to be twice the electromagnetic volume contribution
\bdm\frac{q^2}{2r}$. Discussing geometrical terms, one may wonder
whether a curvature term \bdm\tilde{m}\!\oint \! R\mdg\dS$ makes sense
as well. We will return to this option later on, when confronting the
problem of mass.

\section{Threefold Family Structure}
The most naive interpretation of $\Lambda$ has to do with the
vacuum-expectation-value (VEV) of some world-manifold scalar potential.
Such a familiar point of view (see e.g. inflationary cosmology) has one
immediate advantage. Namely, the scalar potential may admit several
isolated minima. This may open the door for a \underline{finite} family
structure, establishing a viable alternative to Dirac's radial
excitation idea. Our simplest choice is a \underline{real}, that is,
\Ug-neutral, scalar field, with
\beq
S^{(3)}=\oint\mdg\left(g^{\al\beta}\Phi,_{\al}\Phi,_{\beta}+V(\Phi)
\right) \dS\hsp,
\eeq
replacing $\bar{S}^{(3)}$. The reason for choosing a real scalar is
first of all technical. Since the covariant and the ordinary derivatives
are the one and the same in this case, the $a_{\al}$-equation of motion
need not be re-examined. A single complex field, on the other hand,
would not have allowed (a fact soon to be fully appreciated) for a
solution without violating spherical symmetry.

The fact that a real scalar is involved has here a deep theoretical
consequence. As is well known, based on renormalizability arguments, the
bare scalar potential in $d$-dimensions can be at most of order
$\frac{2d}{d-2}$.  With this in mind, we are after the exciting
possibility that a $6^{th}$-order potential plays on our 3-dimensional
world-manifold a role similar to the one played by the ($4^{th}$-order)
Higgs potential in the background space-time. \ Now comes a simple
observation: for a real $\Phi$ we may have, as advertised at most 3
independent isolated minima. For a complex $\Phi$, however, in the
absence of odd self-interaction terms, one \Ug--degenerate minimum may
be solely accompanied by a trivial minimum at the origin. \ Denoting
the three solutions of \bdm\frac{\prt V}{\prt\Phi}=0$ by
$\langle\Phi\rangle_i=\Lambda_i$, each family is characterized by a
different spontaneously induced surface-tension $\Lambda_i$. This in
turn gets directly translated into the Dirac language [3] of different
radii and masses. For example,
\bie
\hf r_i=\frac{3q^2}{4\Lambda_i}\hfc M_i=\frac{9q^2}{4r_i}\hf $\ \ in the
$m\rightarrow\infty$ Dirac limit.\\
Also note that, reflecting the polynomial structure of the potential,
the individual `lepton numbers' are preserved up to exponentially
suppressed tunneling effects.

\section{SU(2)$_{\rm spin}\otimes$U(1)$_{\mg}$ Group Theory}
A magnetic monopole-like configuration is known to have some dramatic
geometrical consequences [12], in particular if it lives on the $S^2$
sphere. To analyze the group theoretical spectrum of the allowed
harmonic excitations, consider a generic secondary-charged
world-manifold scalar field $\psi(y^{\al})$ whose covariant derivative
(with respect to the curved geometry as well as the \Ug\ gauge group) is
given by
\bed
D_{\al}\psi=\psi_{;\al}+iea_{\al}\psi\hsp.
\eed
The corresponding Laplacian is explicitly given by
\beq
g^{\al\beta}D_{\al}D_{\beta}\psi
\!=\!\!\left(\!-\PD{\tau}{2}\!+\!\frac{1}{r^2\sin^2\theta}\!\cdot\!
\PD{\theta}{}\!\left(\!\sin\theta\PD{\theta}{}\!\right)\!\!+\!
\frac{1}{r^2\sin^2\theta} \!\left(\PD{\phi}{}
\!+\!ie\g(1\!-\!\cos\theta)\!\right)^{\! 2}\right)\!\psi\hsp.
\eeq
Up to the additive term \bdm \frac{e^2\g^2}{r^2}$, the angular part of
the Laplacian is recognized as the Casimir operator
$J^2=J^2_1+J^2_2+J^2_3$ of {\sf spatial} $SU(2)$. Indeed, the three
generators are
\bie
J_1=i\sin\phi\PD{\theta}{}+\cos\phi\left(i\frac{\cos\theta}{\sin\theta}
\cdot\PD{\phi}{}+\frac{e\g}{\sin\theta}(1-\cos\theta)\right)\hsp,
\nie
J_2=-i\cos\phi\PD{\theta}{}+
\sin\phi\left(i\frac{\cos\theta}{\sin\theta}\cdot \PD{\phi}{}
+\frac{e\g}{\sin\theta}(1-\cos\theta)\right)\hsp,\prqn 
\nie
J_3=-i\PD{\phi}{}+e\g\hsp.
\eie
An eigenstate of $J_3$, namely, $J_3\psi=s\psi$, which happens to be the
highest weight in its representation, that is $J^+\psi=0$, is found to
be
\beq
\psi\approx\left(\sin\frac{\theta}{2}\right)^{s-e\mg}
\left(\cos\frac{\theta}{2}\right)^{s+e\mg}e^{i(s-e\mg)\phi}\hsp.
\eeq
The lower states of the associated representation can be constructed by
successive $J^-$ operations, with the characteristic effect of
decreasing the powers of $(\sin\frac{\theta}{2})$ and/or
$(\cos\frac{\theta}{2})$ by units of 1 at each step. This tells us that
the representation is non-singular if $s\pm e\g$ is non-negative, and is
furthermore finite provided
\beq
s\pm e\g=\mbox{non-negative integer}\hsp.
\eeq
This completes the classification of the allowed
$SU(2)_{\rm spin}\otimes U(1)_{\mg}$ representations. In particular, a
spin--0 field can only tolerate $e\g=0$. This constitutes our previous
insight regarding the reality of the scalar field $\Phi$ invoked to
resolve the family puzzle. More interesting, however, and soon to be put
to work, are the two independent spin--1/2
representations associated with $e\g=+\frac{1}{2}$ and
$e\g=-\frac{1}{2}$, respectively. They forcefully, demonstrate how a
half--integer spin can spontaneously arise from a sophisticated
apparently scalar configuration. This highly reminds us of the Skyrme
soliton [13], where a lump-like solution of a classical scalar field is
quantized as a fermion. We find it ironic that the famous Dirac
quantization condition $e\g=\frac{n}{2}$ gets realizes here for the
secondary \Ug\ and not, as was originally meant [10], for the parent
\Ue. \ At this point, it is worth recalling once again that g also
serves as the bubble's electric charge $q$.

\section{A Pedagogical No--Go Theorem}
Picking up one of the two spin--1/2 representations, let us study the
role of an additional term:
\beq
\bar{S}^{(4)}=
\oint\mdg\sum_{i=\dna, \upa}
g^{\al\beta}(D_{\al}\psi_i)^{\dagger}(D_{\beta}\psi_i)\dS\hsp,
\eeq
where $D_{\al}\psi_i=(\prt_{\al}+iea_{\al})\psi_i$\hsp. A scalar
potential $W(\psi^{\dagger}\psi)$ is optional, but will distract us from
the main stream. \ As far as the variation with respect to $\psi_i$ is
concerned, two solutions are possible:
\bea
\g=+\frac{1}{2e}\hspace{8mm}\Rightarrow\hspace{8mm}\cases{
\psi_{ \upa}=a\cos\frac{\theta}{2}\hsp,\cb
\psi_{\dna}=b\sin\frac{\theta}{2}e^{-i\phi}\hsp,\cr}
\nie
\nie
\g=-\frac{1}{2e}\hspace{8mm}\Rightarrow\hspace{8mm}\cases{
\psi_{ \upa}= c\sin\frac{\theta}{2}e^{i\phi}\hsp,\cb
\psi_{\dna}=-d\cos\frac{\theta}{2}\hsp.\cr}
\eea
The good news are that the constant of integration g is no longer
arbitrary. Namely, it may only take the specific values
$\pm\frac{1}{2e}$\hsp, corresponding to fixed (opposite) electric
charges. We note in passing that the fractional quark charges are
expected to arise naturally once \Ue\ is embedded within a grand
unifying theory. A unified generalization of the present idea probably
involves a secondary `t Hooft--Polyakov monopole [14].

The above solution may give us the false impression that two scalar
components suffice to describe an electron (or a positron). The
trouble is simple: the above is not a self-consistent solution. To see
the point, concentrate on \bdm\frac{\delta L}{\delta a_{\al}}$ which
receives here its only non-trivial contribution. Self-consistency then
takes the covariant current conservation form:
\bed
\left(\psi^{\dagger}\prt_{\al}\psi-
(\prt_{\al}\psi^{\dagger})\psi\right)
+2iea_{\al}\psi^{\dagger}\psi=0\hsp,
\eed
which is automatically satisfied for $\al=\tau$, $\theta$, but fails to
hold for $\al=\phi$, unless
\bed
a^2=b^2\hfc c^2=d^2\hsp,
\eed
but this is sick from the $SU(2)_{\rm spin}$ point of view (e.g.
$J^{\pm}\psi$ takes us off the representation). However, such a
spontaneous symmetry violation comes with no surprise. Having in
mind that the Dirac algebra does not have a $2\times2$ Pauli realization
(The $\gamma^{\mu}$--matrices are $4\times4$ in their minimal
representation), one could have expected some kind of no-go situation.
Indeed, the remedy relies upon constructive co-operation between the two
independent spin-1/2 representations.

\section{The Electron (Positron) Configuration}
Taking into account both available spin--1/2 representations,
$\bar{S}^{(4)}$ is replaced by the charge--conjugation invariant
($e\rightarrow -e$) kinetic term
\beq
S^{(4)}=\oint\mdg\left(\sum_{i=+,-\atop j=\upa,\dna}g^{\al\beta}
(D_{\al}\psi_{ij})^{\dagger}(D_{\beta}\psi_{ij})+W(\psi^{\dagger}\psi)
\right)\dS\hsp,
\eeq
where $D_{\al}\psi_{\pm}=(\prt_{\al}\pm iea_{\al})\psi_{\pm}$\ .\
The corresponding self-consistency condition takes the generalized form
\beq
(\psi^{\dagger}_+\prt_{\al}\psi_+-(\prt_{\al}\psi^{\dagger}_+) \psi_+)
-(\psi^{\dagger}_-\prt_{\al}\psi_--( \prt_{\al}\psi^{\dagger}_-)\psi_-)
+2iea_{\al}(\psi^{\dagger}_+\psi_++\psi^{\dagger}_-\psi_-)=0\hsp.
\eeq
At this stage, our discussion bifurcates. First we present the electron
configuration ($\g=+\frac{1}{2e}$), and then go on to consider its
positron ($\g=-\frac{1}{2e}$) companion.

{\bf The electron:} For $\g=+\frac{1}{2e}$, we derive
\beq
\Psi=\cases{\psi_{+\upa}=a\cos\frac{\theta}{2}\cb
            \psi_{+\dna}=b\sin\frac{\theta}{2}e^{-i\phi}\cb
            \psi_{-\upa}=c\sin\frac{\theta}{2}e^{i\phi}\cb
            \psi_{-\dna}=-d\cos\frac{\theta}{2}\cr}
\eeq
subject to \bdm\hff a^2+d^2=b^2+c^2\hsp,\EQ{17}
by virtue of the above self-consistency condition. \ Dismissing the two
redundant solutions ($a=b=0$, and $c=d=0$) discussed earlier, we can
finally identify the polarized electron configurations
\beq
\label{18}
\Psi_{\upa}=\cases{\psi_{+\upa}=\A\cos\frac{\theta}{2}\cr
            \psi_{+\dna}=0\cr
            \psi_{-\upa}=\A\sin\frac{\theta}{2}e^{i\phi}\cr
            \psi_{-\dna}=0}\hfc
\psi_{\dna}=\cases{\psi_{+\upa}=0\cr
            \psi_{+\dna}=\A\sin\frac{\theta}{2}e^{-i\phi}\cr
            \psi_{-\upa}=0\cr
            \psi_{-\dna}=-\A\cos\frac{\theta}{2}\cr}\hsp.
\eeq
The underlying spherical symmetry is expressed by the fact that, for
each configuration, $\Psi^{\dagger}\Psi$ is independent of $\phi$,
$\theta$. The role of the scalar potential $W(\Psi^{\dagger}\Psi)$ in
fixing $\A(r,t)$ is to be clarified soon. \ As we shall see, its
presence is absolutely necessary for having a static ($t$-independent)
solution. The $6^{th}$-order polynomial structure of
$W(\Psi^{\dagger}\Psi)$ and the complexity of $\Psi$ indicate that $\A$
can take at most \underline{one} non-vanishing value. The total number
of (three) families is not expected to increase by the introduction of
$\psi_{ij}$.

{\bf The Positron:} For $\g=-\frac{1}{2e}$\ , interchanging the roles
played by $\psi_+$ and $\psi_-$, a similar calculation reveals
\beq
\label{19}
\Psi^c=\cases{
\psi_{+\upa}=a\sin\frac{\theta}{2}e^{i\phi}\cb
\psi_{+\dna}=-b\cos\frac{\theta}{2}\cb
\psi_{-\upa}=c\cos\frac{\theta}{2}\cb
\psi_{-\dna}=d\sin\frac{\theta}{2}e^{-i\phi}\cr}\hsp,
\eeq
again with \bdm\hf a^2+d^2=b^2+c^2$\ . \ The physical polarized
positron configurations are the following
\beq
\label{20}
\Psi^c_{\upa}=\cases{
\psi_{+\upa}=\A\sin\frac{\theta}{2}e^{i\phi}\cr
\psi_{+\dna}=0\cr
\psi_{-\upa}=\A\cos\frac{\theta}{2}\cr
\psi_{-\dna}=0\cr}\hfc
\Psi^c_{\dna}=\cases{
\psi_{+\upa}=0\cr
\psi_{+\dna}=-\A\cos\frac{\theta}{2}\cr
\psi_{-\upa}=0\cr
\psi_{-\dna}=\A\sin\frac{\theta}{2}e^{-i\phi}\cr}\hsp.
\eeq

\section{Light Mass?}
Recalling that the full action $S^{(1)}+S^{(2)}+S^{(3)}+S^{(4)}$
exhibits no explicit $t(y^{\al})$ dependence, the total energy of the
system is clearly conserved. To be somewhat more specific, with
spherical symmetry imposed, we have
\bed
\PD{\tau}{}\left(\frac{\delta L}{\delta t,_{\tau}}+
\frac{q^2}{2r}\right)=0\hsp,
\eed
where the $\frac{q^2}{2r}$ piece is the Dirac contribution of the
surrounding electromagnetic field. The $t,_{\al}$--dependence of the
Lagrangian is hidden within
\bdm\frac{\delta g_{\tau\tau}}{\delta t,_{\tau}}=-2$ , \bdm\frac{\delta
A_{\tau}}{\delta t,_{\tau}}=\frac{q}{r}$. The general analysis of the
total energy lies beyond the scope of the present paper. Here, our only
intention is to demonstrate the existence of stable static
configurations. We minimize $E(r,\Phi,\A)$, and speculate about the
lightness  of the mass.

It takes some algebra to arrive at the energy formula
\beq
E(r,\Phi,\A)=(V(\Phi)+W(\A^2))r^2+\frac{1}{2}\A^2+\frac{\g^2}{2mr^2}+
\frac{3q^2}{2r}\hsp,
\eeq
where \bdm W_{\rm eff}\equiv W+\frac{\A^2}{2r^2}$ is the fermionic
contribution to the surface tension . $V(\Phi)+W(\A^2)$ can be harmlessly
embedded within a more general potential $U(\Phi,\A^2)$. Noticeably, a
major Dirac drawback is still present. For radii smaller than
$10^{-15}$cm, one obtains masses heavier than $100m_e$. Is there a
physical way to make the mass lighter, or perhaps arbitrarily light?
\ To answer this question, we recall that the previously mentioned
curvature option has not yet been fully exploited. The bubble curvature
may enter either via a conformal coupling \bdm -\frac{1}{8}\oint R\Phi^2
\mdg\dS$, which gives $\Delta E=+\frac{1}{8}\Phi^2$, and/or via `surface
gravity':
\beq
S^{(5)}=\tilde{m}\oint R\mdg\dS\hsp,
\eeq
whose tenable contribution $\Delta E=-2\tilde{m}$ is of-course most
welcome. This may be another indication that an Einstein--Maxwell
generalization is mandatory, with the self-consistent mass $M$
parametrizing the surrounding Reissner--Nordstrom geometry. At any rate,
the overall situation reminds us of the standard electroweak model,
where the origin of mass is well established yet nothing specific can be
said about individual fermionic masses.

\section{Summary}
Invoking world-manifold gauge field theory, we have revived the Dirac
idea of a finite size electron, \underline{spin} and \underline{family}
included. Our final product is a threefold family of electrically
charged spin--1/2 static Dirac bubbles. The main result, formulated by
eqs. \eref{18}, \eref{20}, establishes a one-to-one correspondence
between four world-manifold scalar configurations and the four
components of the Dirac spinor. On the technical side, a key role is
played by the spherically symmetric magnetic monopole ground-state
configuration of the secondary \Ug\ gauge field. Namely, it provides the
locally conserved charge distribution for the primary \Ue\ electric
monopole, and furthermore allows (in a group theoretical manner) for the
spontaneous generation of spin. The strength of the magnetic monopole,
also serving as the bubble's electric charge, is uniquely fixed by the
Dirac quantization condition. The family puzzle is resolved by
minimizing a ($6^{th}$-order) surface-tension potential; at most three
families are expected. As far as the Fermi mass problem is concerned, no
quantitative progress can be reported at the present stage. The
situation may change, however, once a better understanding of the
so-called surface gravity becomes available. The complement list of
topics to be discussed in future includes the interactions with external
sources, the non-Abelian generalization
(focusing on $SU(3)_{\rm color}\otimes U(1)_{\rm e.m.}$), the
self-consistent gravitational treatment, and the provocative massless
limit.

\immediate\write16{***********************************************}
\immediate\write16{* and DON'T forget to LaTeX this file TWICE ! *}
\immediate\write16{***********************************************}
\end{document}